\documentstyle[12pt]{article}
\def\be{\begin{eqnarray}}
\def\ee{\end{eqnarray}}
\def\eq{\label}

\def\abstract#1{\vskip 7mm 
\begin{center}{\large Abstract}\par \bigskip
\begin{minipage}[c]{12cm}
\small #1
\end{minipage}
\end{center}
}
\def\title#1{\begin{center}{\Large\bf #1}\end{center}}
\def\author#1{\vskip 5mm \begin{center}{#1}\end{center}}
\def\address#1{\begin{center}{\it #1}\end{center}}
\newcommand{\bfr}{\begin{flushright}}
\newcommand{\efr}{\end{flushright}}
\begin{document}
\vspace*{-0cm}
\title{Infinite Number of Stationary Soliton Solutions to Five-dimensional Vacuum Einstein Equation}
\author{Takahiro AZUMA\footnote{E-mail: azuma@dokkyo.ac.jp} 
and Takao KOIKAWA\footnote{E-mail: koikawa@otsuma.ac.jp}
}
\vspace{1cm}
\address{
${}^1$ Department of Languages and Culture, Dokkyo University, \\ 
Soka 340-0042, Japan \\
${}^2$ School of Social Information Studies, Otsuma Women's University, \\
Tama 206-0035, Japan
}
\vspace{3.5cm}
\abstract{
We obtain an infinite number of soliton solutions to the the five-dimensional stationary Einstein equation with axial symmetry by using the inverse scattering method. \ We start with the five-dimensional Minkowski space as a seed metric to obtain these solutions. \ The solutions are characterized by two soliton numbers and a constant appearing in the normalization factor related to a coordinate condition. \ We show that the (2,0)-soliton solution is identical to the Myers-Perry solution with one angular momentum by imposing a condition between parameters. \ We also show that the (2,2)-soliton solution is different from the black ring solution discovered by Emparan and Reall, although one component of the metric of two metrics can be identical.}

\newpage
\setcounter{page}{2}

Black hole solutions in the five-dimensional space-time have been attracted considerable attention. \ It has been shown that the structure of solution in five dimensions is much more complicated than that in four dimensions. \ The Myers-Perry rotating black hole solution\cite{MyersPerry} has horizon topology $S^3$ while the rotating black ring solution discovered by Emparan and Reall\cite{blackring} has horizon topology $S^2 \times S^1$. \ These two solutions are connected with each other. \ It is known that the general black ring solution is reduced to the Myers-Perry black hole solution with one angular momentum at a certain limit\cite{harmark}. \ The general black ring solution generically has conical singularities. \ These singularities can be avoided by choosing the periods of some coordinates appropriately in the stationary case, while they cannot be avoided in the static case. 

Recently one of the authors\cite{K} used the soliton technique, which has been known to construct an infinite number of stationary and axisymmetric solutions to the Einstein equation in four dimensions, to obtain an infinite number of static solutions in five dimensions. \ The solutions are characterized by two integers representing the soliton numbers. \ He showed that the first non-trivial example of these solutions coincides with the static black ring solution. \ Mishima and Iguchi\cite{M-I} also used a solitonic solution-generating technique to obtain axisymmetric stationary solutions in five dimensions. \ They showed that their solutions include the black ring solutions that are different from the solution discovered by Emparan and Reall. 

In this paper we apply the inverse scattering method\cite{Bel-Zak}, known as one of soliton techniques, to the five-dimensional stationary and axisymmetric Einstein equation. \ This method is expected to provide us with an infinite series of solutions in five dimensions, which may include not only the known rotating black hole and black ring solutions but also new type of solutions in five dimensions.

We start with the five-dimensional stationary and axisymmetric metric in the canonical coordinates. \ The Einstein equation is divided into a differential equation for a $3 \times 3$ matrix and equations which are integrable by using the solution to the matrix equation. \ In contrast to the static case where the $3 \times 3$ matrix is diagonal, the matrix has non-zero off-diagonal entities in the stationary case. \ Since it is difficult to apply the inverse scattering method to the equation for the $3 \times 3$ matrix with full entities, we decompose it into a $2 \times 2$ matrix and a scalar component. \ This means that we treat the five-dimensional stationary solutions with one angular momentum. \ The solutions thus obtained are characterized by two integers corresponding to the soliton numbers of solutions for the matrix and scalar component. \ In addition to these integers the solutions are characterized by a real number appearing in the  normalization factor related to a coordinate condition. \ We find that the (2,0)-soliton solution becomes the Myers-Perry solution with one angular momentum when we make an appropriate choice for constants in the solution. \ We also find that the (2,0)-soliton or (2-2)-soliton solution is not identical to the general black ring solution although the scalar component coincides with the corresponding component of the black ring solution.

The metric is given by
\be
ds^2=f(d\rho^2+dz^2)+g_{ab}dx^adx^b,
\ee
where $f$ and $g_{ab}$ are functions of $\rho$ and $z$, and the suffices of $a$ and $b$ of $g_{ab}$ run from 0 to 2. \ The five-dimensional Einstein equation reads
\be
&&\left(\rho G,_{\rho}G^{-1}\right),_{\rho}+\left(\rho G,_{z}G^{-1}\right),_{z}=0,\eq{eq:G}\\
&&(\ln f),_{\rho}=-\frac{1}{\rho}+\frac{1}{4\rho}Tr(U^2-V^2),
\eq{eq:f1}\\
&&(\ln f),_{z}=\frac{1}{2\rho}Tr(UV),\eq{eq:f2}
\ee
where $G$ is a $3\times 3$ matrix given by $G=(g_{ab})$, and $U$ and $V$ are matrices defined by
\be
U=\rho G,_{\rho}G^{-1},\quad V=\rho G,_{z}G^{-1}.
\ee
The coordinate condition that is compatible with the equations (\ref{eq:G})--(\ref{eq:f2}) is
\be
\det G=-\rho^2.\eq{eq:condition1}
\ee
In order to obtain soliton solutions to the five-dimensional Einstein equation we decompose the matrix $G$ into a $2\times 2$ matrix $g$ defined by
\be
g=\pmatrix{g_{00} & g_{01} \cr
           g_{01} & g_{11}}, 
\ee
and the scalar component $g_{22}$. \ Then, the equation (\ref{eq:G}) is decomposed into two equations
\be
&&\left(\rho g,_{\rho}g^{-1}\right),_{\rho}+\left(\rho g,_{z}g^{-1}\right),_{z}=0,\eq{eq:g}\\
&&\left[\rho(\ln g_{22}),_{\rho}\right],_{\rho}
+\left[\rho (\ln g_{22}),_{z}\right],_{z}=0,\eq{eq:g22}
\ee
and we have the relations
\be
&&\frac{1}{4\rho}Tr(U^2-V^2)=\frac{1}{4\rho}Tr(u^2-v^2)
+\frac{\rho}{4}\left((\ln g_{22}),_{\rho}^2-(\ln g_{22}),_z^2\right),
\\
&&\frac{1}{2\rho}Tr(UV)=\frac{1}{2\rho}Tr(uv)+\frac{\rho}{2}
(\ln g_{22}),_{\rho}(\ln g_{22}),_z,
\ee
where
\be
u=\rho g,_{\rho}g^{-1},\quad v=\rho g,_{z}g^{-1}.
\ee
The coordinate condition (\ref{eq:condition1}) becomes
\be
(\det g)g_{22}=-\rho^2.\eq{eq:condition2}
\ee
We apply the inverse scattering method to both the equations (\ref{eq:g}) and (\ref{eq:g22}). \ The seed metric from which we generate soliton solutions is the five-dimensional Minkowski metric
\be
ds^2=\frac{1}{2\sqrt{\rho^2+z^2}}(d\rho^2+dz^2)-(dx^0)^2+\mu^*(dx^1)^2-\mu(dx^2)^2,
\eq{eq:Minkowski}
\ee
where
\be
\mu^*=-z+\sqrt{\rho^2+z^2},\\
\mu=-z-\sqrt{\rho^2+z^2}.
\ee
The obtained solutions for $g$ and $g_{22}$ do not generally satisfy the condition (\ref{eq:condition2}), and therefore we introduce a renormalization factor $N$ to get physical solutions by
\be
g^{(ph)}=N^{\frac{\gamma}{2}}g, \quad g_{22}^{(ph)}=N^{1-\gamma}g_{22},
\eq{eq:physical_solution}
\ee
where $\gamma$ is an arbitrary real constant. \ Using the obtained solutions,  $N$ is written as
\be
N=-\frac{\rho^2}{(\det g )g_{22}}.
\ee

We first give the soliton solutions to the equation (\ref{eq:g22}). \ The seed metric for $g_{22}$ is $(g_{22})_0=-\mu$. \ Introducing a function $h$ by
\be
g_{22}=-\mu h,
\ee
we obtain the soliton solution
\be
h=\prod_{k=1}^n(i\frac{\tilde\mu_k}{\rho}),
\ee
where
\be
\tilde\mu_k=\tilde w_k-z +(-1)^k \sqrt{\rho^2+(\tilde w_k-z)^2}.\quad (k=1,2,\cdots,n)
\ee
In the above solution $n$ is an integer number which is called soliton number and $\tilde w_k$ are arbitrary constants. \ We next solve the matrix equation (\ref{eq:g}) with the seed metric
\be
g_0=\pmatrix{-1 & 0 \cr
           0 & \mu^*}.
\ee
The so-called wave matrix $\psi(\lambda)$ in the inverse scattering method\cite{Bel-Zak}, which coincides with $g_0$ when the parameter $\lambda=0$, is given in this case by
\be
\psi(\lambda)=\pmatrix{-1 & 0 \cr
           0 & \mu^*-\lambda}.
\ee
In order to construct the soliton solution for $g$ by using this matrix function we introduce constant vectors by
\be
\vec m_0^{(k)}=\pmatrix{p_k \cr q_k},\quad (k=1,2,\cdots,m)
\ee
and define two-vectors by
\be
\vec m^{(k)}=\psi^{-1}(\mu_k)\vec m_{0}^{(k)}
=\pmatrix{-p_k \cr q_k(\mu^*-\mu_k)^{-1}},\quad (k=1,2,\cdots,m)
\ee
with
\be
\mu_k=w_k-z +(-1)^k \sqrt{\rho^2+(w_k-z)^2},\quad (k=1,2,\cdots,m)
\ee
where $w_k$ are constants. \ We also define the matrix $\Gamma$ whose components are given by
\be
\Gamma_{kl}=\frac{\vec m^{(k)t}g_0\vec m^{(l)}}{(\rho^2+\mu_k \mu_k)}.
\quad (k,l=1,2,\cdots,m)
\ee
Their explicit expressions are
\be
\Gamma_{kl}=\frac{-p_kp_l(\mu^*-\mu_k)(\mu^*-\mu_l)+q_kq_l\mu^*}
{(\mu^*-\mu_k)(\mu^*-\mu_l)(\rho^2+\mu_k\mu_l)}.\quad (k,l=1,2,\cdots,m)
\ee
We now obtain the soliton solution
\be
g_{ab}=(g_0)_{ab}
-\sum_{k,l}^m\mu_k^{-1}\mu_l^{-1}N_a^{(k)}(\Gamma^{-1})_{kl}N_b^{(l)},
\quad (a,b=0,1)
\ee
where $N_a^{(k)}$ are the components of vectors which are defined by
\be
\vec N^{(k)}=g_0 \vec m^{(k)}=\pmatrix{p_k \cr q_k\mu^*(\mu^*-\mu_k)^{-1}}.\quad (k=1,2,\cdots,m)
\ee
In the above solution the integer number $m$ is the soliton number for the solution $g$. \ We have given the solutions for $g$ and $g_{22}$ that are characterized by the soliton numbers $m$ and $n$, respectively. \ We call these solutions $(m,n)$-soliton solution to the five-dimensional Einstein equation. \ By substituting this $(m,n)$-soliton solution into the equation (\ref{eq:f1}) and (\ref{eq:f2}), we integrate the equation directly to obtain
\be
f&=&C_{mn}\frac{\rho^{m+n^2/4+1}}{\mu^{n/2}\sqrt{\mu^2+\rho^2}\sqrt{\mu^{*2}+\rho^2}}
\prod_{k=1}^m\frac{\mu_k^2}{\mu_k^2+\rho^2}\det\Gamma \cr
&\times&\prod_{k=1}^n(\tilde\mu_k-\mu)\prod_{k>l}^n(\tilde\mu_k-\tilde\mu_l)
\prod_{k=1}^n[\tilde\mu_k^{n-2}(\tilde\mu_k^2+\rho^2)]^{-1/2},
\ee
where $C_{mn}$ is a constant to be determined by the asymptotic flat condition.

In order to compare the obtained soliton solution with the known five-dimensional rotating black hole or black ring solutions, we consider the (2,0)-soliton or (2,2)-soliton solution and the case $\tilde\mu_k=\mu_k$, i.e. the case $\tilde w_k=w_k$. \ The components of the 2-soliton solution for $g$ are explicitly given by
\be
g_{00}&=&\rho^2\mu_1^{-2}\mu_2^{-2}Q^{-1} \cr
&\times&
\left( (\mu_2-\mu_1)^2
      [p_1p_2\rho^2(\mu^*-\mu_1)(\mu^*-\mu_2)+q_1q_2\mu_1\mu_2\mu^*]^2 \right. \cr
       && - \left. \mu^*(\rho^2+\mu_1\mu_2)^2 
      [p_1q_2(\mu^*-\mu_1)\mu_2-q_1p_2(\mu^*-\mu_2)\mu_1]^2
\right),\\
g_{01}&=&(\mu_2-\mu_1)\mu^*\rho^2(\rho^2+\mu_1\mu_2)\mu_1^{-2}\mu_2^{-2}Q^{-1} \cr
&\times&
\left( p_1q_1(\mu_1^2+\rho^2)\mu_2(\mu^*-\mu_1)[p_2^2(\mu^*-\mu_2)^2-q_2^2\mu^*]
\right. \cr
    &&
\left. -p_2q_2(\mu_2^2+\rho^2)\mu_1(\mu^*-\mu_2)[p_1^2(\mu^*-\mu_1)^2-q_1^2\mu^*]
\right),\\     
g_{11}&=&-\mu^*\rho^2\mu_1^{-2}\mu_2^{-2}Q^{-1} \cr
&\times&
\left((\mu_2-\mu_1)^2
      [p_1p_2\mu_1\mu_2(\mu^*-\mu_1)(\mu^*-\mu_2)+q_1q_2\rho^2\mu^*]^2 \right. \cr
      && - \left. \mu^*(\rho^2+\mu_1\mu_2)^2
      [p_1q_2(\mu^*-\mu_1)\mu_1-q_1p_2(\mu^*-\mu_2)\mu_2]^2
\right),
\ee
where
\be
Q&=&-
\rho^2(\mu_2-\mu_1)^2
      [p_1p_2(\mu^*-\mu_1)(\mu^*-\mu_2)-q_1q_2\mu^*]^2 \cr
      &&-\mu^*(\rho^2+\mu_1\mu_2)^2
      [p_1q_2(\mu^*-\mu_1)-q_1p_2(\mu^*-\mu_2)]^2.
\eq{eq:D1}
\ee
These components give us the determinant of $g$
\be
\det g=-\frac{\rho^4 \mu^*}{\mu_1^2 \mu_2^2},
\ee
and then the normalization factor $N$ becomes
\be
N=\frac{\mu_1^2\mu_2^2}{\rho^4 h}.
\ee

We first study the (2,0)-soliton solution. \ Introducing two constants $\sigma$ and $z_0$ instead of $w_1$ and $w_2$ in $\mu_1$ and $\mu_2$ through the relations $w_1=z_0-\sigma$ and $w_2=z_0+\sigma$ and transforming the variable $z$ to $z+z_0$, we have
\be
\mu_1&=&-(z+\sigma)-\sqrt{\rho^2+(z+\sigma)^2}, \eq{eq:mu1} \\
\mu_2&=&-(z-\sigma)+\sqrt{\rho^2+(z-\sigma)^2}, \eq{eq:mu2} \\
\mu&=&-(z+z_0)-\sqrt{\rho^2+(z+z_0)^2}, \eq{eq:mu} \\
\mu^*&=&-(z+z_0)+\sqrt{\rho^2+(z+z_0)^2}. \eq{eq:mus}
\ee 
For the 0-soliton solution for $g_{22}$ we have $h=1$ and
\be
N=\frac{\mu_1^2\mu_2^2}{\rho^4}.
\ee
When we choose the constant $\gamma$ in (\ref{eq:physical_solution}) such that $\gamma=1$, we obtain the physical solutions
\be
g^{(ph)}=-\frac{\mu_1\mu_2}{\rho^2}g, \quad g_{22}^{(ph)}=-\mu.
\eq{eq:physical_solution2}
\ee
By substituting these physical solutions instead of $g$ and $g_{22}$ into the equation (\ref{eq:f1}) and (\ref{eq:f2}), we integrate the equation to obtain the physical solution for $f$:
\be
f^{(ph)}=\frac{\tilde C_{20}\rho Q}
{\mu^*\sqrt{\mu^2+\rho^2}\sqrt{\mu^{*2}+\rho^2}(\mu^*-\mu_1)(\mu^*-\mu_2)
(\mu_1^2+\rho^2)(\mu_2^2+\rho^2)},
\eq{eq:physical_f}
\ee
where $\tilde C_{20}$ is a constant and $Q$ is given in (\ref{eq:D1}).
\ The form of $g_{22}^{(ph)}$ in (\ref{eq:physical_solution2}) is the same as that of the corresponding component of the Myers-Perry solution with one angular momentum. \ However, in the asymptotic region where $\sqrt{\rho^2+z^2}\to\infty$ with $z/\sqrt{\rho^2+z^2}$ finite, $g_{01}^{(ph)}$ has the form
\be
g_{01}^{(ph)}=-2 c_0+{\cal O}((\rho^2+z^2)^{-1/2}),
\ee
where $c_0$ is a constant given by
\be
c_0=\frac{2p_1q_2\sigma}{2p_1p_2(\sigma+z_0)+q_1q_2},
\eq{eq:c0}
\ee
and therefore we find that the obtained solution (\ref{eq:physical_solution2}) does not approach to the five-dimensional Minkowski metric (\ref{eq:Minkowski}). \ In order to get an asymptotically flat solution we introduce a new time variable $t$ by the coordinate transformation
\be
t=x^0+2c_0x^1.
\eq{eq:coordinate_tr}
\ee
Then, in the new coordinates we have an asymptotically flat solution
\be
g_{tt}&=&g_{00}^{(ph)}, \eq{eq:gtt} \\
g_{t\phi}&=&g_{01}^{(ph)}-2c_0g_{00}^{(ph)}, \eq{eq:gtphi} \\
g_{\phi\phi}&=&g_{11}^{(ph)}-4c_0g_{01}^{(ph)}+4c_0^2g_{00}^{(ph)}, \eq{eq:gphiphi} \\
g_{\psi\psi}&=&g_{22}^{(ph)}, \eq{eq:gpsipsi}
\ee
where we have put $x^1=\phi$ and $x^2=\psi$. \ The condition that $f^{(ph)}$ should behave as $1/(2\sqrt{\rho^2+z^2})$ in the asymptotic region determines the constant $\tilde C_{20}$:
\be 
\tilde C_{20}=\frac{2(\sigma+z_0)}{[2p_1p_2(\sigma+z_0)+q_1q_2]^2}.
\ee

We now clarify the relation between the present solution and the Myers-Perry solution with one angular momentum. \ In the above solution, the constants $z_0$ and $\sigma$ are independent and therefore we choose $z_0$ such that $z_0=-\sigma$. \ In this case, $\mu^*=\mu_2$, and then we find
\be
g_{tt}&=&-\frac{\mu_2[(\rho^2+\mu_1\mu_2)^2p_1^2-\mu_1^2\mu_2q_1^2]}
{\mu_1[(\rho^2+\mu_1\mu_2)^2p_1^2+\rho^2\mu_2q_1^2]},\\
g_{t\phi}&=&\frac{\mu_2p_1}{\mu_1q_1[(\rho^2+\mu_1\mu_2)^2p_1^2+\rho^2\mu_2q_1^2]}
\cr
&\times&\left(4\sigma(\rho^2+\mu_1\mu_2)^2p_1^2-
[(\rho^2+\mu_1^2)(\rho^2+\mu_1\mu_2)+4\mu_1^2\mu_2\sigma]q_1^2\right),\\
g_{\phi\phi}&=&-\frac{1}{\mu_1[(\rho^2+\mu_1\mu_2)^2p_1^2+\rho^2\mu_2q_1^2]q_1^2}
\cr
&\times&\left(16\mu_2(\rho^2+\mu_1\mu_2)^2\sigma^2p_1^4+\mu_2\rho^4q_1^4
-[\mu_1^2(\rho^2+\mu_1\mu_2)^2\right.\cr
&&+\left. 8\mu_2(\rho^2+\mu_1^2)(\rho^2+\mu_1\mu_2)\sigma +16\mu_1^2\mu_2^2\sigma^2]p_1^2q_1^2\right),\\
g_{\psi\psi}&=&-\mu,\\
f^{(ph)}&=&\frac{2\rho\sqrt{\rho^2+(z-\sigma)^2}(\mu_2-\mu_1)
[(\rho^2+\mu_1\mu_2)^2p_1^2+\rho^2\mu_2q_1^2]}
{\mu_2\sqrt{\mu^2+\rho^2}(\mu_1^2+\rho^2)(\mu_2^2+\rho^2)^{3/2}q_1^2}.
\ee
If we introduce the variables $r$ and $\theta$ through the relations
\be
\rho=\frac{1}{2}r\sqrt{r^2-r_0^2+a^2}\sin 2\theta,\quad
z=\frac{1}{4}(2r^2-r_0^2+a^2)\cos 2\theta,
\ee
and the constants $r_0$ and $a$ through the relations
\be
\sigma=-\frac{r_0^2-a^2}{4},\quad
\frac{p_1^2}{q_1^2}=\frac{a^2}{r_0^2-a^2},
\ee
we have
\be
\mu&=&-r^2\cos^2\theta,\\ 
\mu_1&=&-(r^2-r_0^2+a^2)\cos^2\theta,\\
\mu_2&=&(r^2-r_0^2+a^2)\sin^2\theta,
\ee 
and then we obtain the Myers-Perry solution with one angular momentum
\be
ds^2&=&-dt^2+r_0^2\Sigma^{-1}(dt-a\sin^2\theta d\phi)^2+(r^2+a^2)\sin^2\theta d\phi^2 \cr
&&+r^2\cos^2\theta d\psi^2+\Sigma(\Delta^{-1}d\rho^2+d\theta^2), 
\ee
where
\be
\Delta =r^2-r_0^2+a^2,\quad \Sigma =r^2+a^2\cos^2\theta.
\ee

We next study the (2,2)-soliton solution. \ In this case $h$ is given by $h=-\mu_1\mu_2/\rho^2$ and therefore $N=-\mu_1\mu_2/\rho^2$. \ If we choose $\gamma$ such that $\gamma=1$, we then obtain the physical solutions
\be
g^{(ph)}=\left(-\frac{\mu_1\mu_2}{\rho^2}\right)^{1/2}g, \quad\quad
g_{22}^{(ph)}=\frac{\mu\mu_1\mu_2}{\rho^2}.
\eq{eq:physical_solution3}
\ee
Note that in the case of the (2,0)-soliton solution with $\gamma=1/2$ we have the physical solution that is identical to (\ref{eq:physical_solution3}). \ For $\mu_1$, $\mu_2$, $\mu$ and $\mu^*$ appearing in the above solution (\ref{eq:physical_solution3}) we use the same expressions as in (\ref{eq:mu1})--(\ref{eq:mus}). \ As in the previous case of the (2,0)-soliton solution with $\gamma=1$, the obtained solution does not approach to the five-dimensional Minkowski metric. \ We need the coordinate transformation (\ref{eq:coordinate_tr}) with the same constant $c_0$ as in (\ref{eq:c0}). \ After this coordinate transformation we obtain the asymptotically flat physical solutions that are expressed in the same forms as in (\ref{eq:gtt})--(\ref{eq:gpsipsi}). 

In the present solution we find that the form of metric component $g_{\psi\psi}$ given by (\ref{eq:gpsipsi}) and (\ref{eq:physical_solution3}) coincides with that of  corresponding component of the general black ring solution. \ In order to study the relation between two solutions in detail, we expand the metric components of both the solutions in the asymptotic region where $\sqrt{\rho^2+z^2}\to\infty$ with $z/\sqrt{\rho^2+z^2}$ finite. \ For our (2,2)-soliton solution we have
\be
g_{tt}&=&-1+\frac{c_1+2c_0^2}{\sqrt{\rho^2+z^2}}+\cdots,\\
g_{t\phi}&=&\left( c_3+c_2-2c_0^3\right)\frac{\sqrt{\rho^2+z^2}-z}{\rho^2+z^2}
+\cdots,\\
g_{\phi\phi}&=&\left(\sqrt{\rho^2+z^2}-z\right)
\left( 1+\frac{c_4+2c_0^2}{\sqrt{\rho^2+z^2}}\right.+\cdots,\\
g_{\psi\psi}&=&\left(\sqrt{\rho^2+z^2}+z\right)
\left( 1+\frac{2\sigma+z_0}{\sqrt{\rho^2+z^2}}\right.+\cdots,
\ee
where
\be
c_1&=&\frac{[6p_1p_2(\sigma+z_0)-q_1q_2]\sigma}{2p_1p_2(\sigma+z_0)+q_1q_2},\\
c_2&=&\frac{-2p_1q_2[2p_1p_2(\sigma+z_0)(3\sigma-z_0)-q_1q_2(3\sigma+z_0)]\sigma}
{[2p_1p_2(\sigma+z_0)+q_1q_2]^2},\\
c_3&=&\frac{2p_2q_1(\sigma+z_0)\sigma}{2p_1p_2(\sigma+z_0)+q_1q_2},\\
c_4&=&\frac{2p_1p_2(\sigma+z_0)(\sigma-z_0)-q_1q_2(3\sigma+z_0)}
{2p_1p_2(\sigma+z_0)+q_1q_2}.
\ee
For the corresponding components of general black ring solution\cite{harmark} they are expanded as
\be
g^{(BR)}_{tt}&=&-1+\frac{2b(1-c)\kappa^2}{1-b}\frac{1}{\sqrt{\rho^2+z^2}}+\cdots,\\
g^{(BR)}_{t\phi}&=&-\sqrt{\frac{2b(1+b)(b-c)}{1-b}}\frac{(1-c)\kappa^3}{1-b}
\frac{\sqrt{\rho^2+z^2}-z}{\rho^2+z^2}
+\cdots,\\
g^{(BR)}_{\phi\phi}&=&\left(\sqrt{\rho^2+z^2}-z\right)
\left[ 1+\frac{(1+b-2c)\kappa^2}{1-b}\frac{1}{\sqrt{\rho^2+z^2}}\right.+\cdots,\\
g^{(BR)}_{\psi\psi}&=&\left(\sqrt{\rho^2+z^2}+z\right)
\left[ 1+\frac{(2c-1)\kappa^2}{\sqrt{\rho^2+z^2}}\right.+\cdots,
\ee
where $b$, $c$ and $\kappa$ are constants. \ We here assume the relations  $\sigma=c\kappa^2$ and $z_0=-\kappa^2$ and then we have $g_{\psi\psi}=g^{(BR)}_{\psi\psi}$. \ Comparing the remaining components we have the relations in which $b$ and $c$ are expressed in terms of $p_1$, $p_2$, $q_1$, $q_2$ and $z_0$. \ However, we find that the relations thus obtained are not compatible with those obtained by comparing the higher-order terms in the expansions of these components. \ Moreover, we find that the component $g_{tt}$ behaves as $g_{tt}\to 0$ in the region $-\sigma<z<\sigma$ on the $z$-axis given by $\rho=0$, while $g^{(BR)}_{tt}$ stays finite in the corresponding region on the axis. \ Therefore we conclude that our (2,2)-soliton solution does not coincide with the general black ring solution.
 
In this paper, we obtained an infinite number of soliton solutions to the five-dimensional stationary Einstein equation with axial symmetry. \ We start with the five-dimensional Minkowski space as a seed metric to obtain these solutions. \ 
The solutions are characterized by two soliton numbers and a constant appearing in the normalization factor related to a coordinate condition. \ We studied the (2,0)- and (2,2)-soliton solutions in order to compare with the known solutions, and we found that the (2,0)-soliton solution with the asymptotic flatness condition is identical to the Myers-Perry solution with one angular momentum by imposing a condition between parameters. \ We also found that the (2,2)-soliton solution is different from the black ring solution, although one component of the metric of two metrics can be identical. \ The detailed structure of the solutions could be clarified by studying the behavior of the solutions near $\rho=0$, or the rod structure.

Here, the matrix $G$ has only two non-zero off-diagonal component. \ We can also discuss the matrix with more non-zero components. \ These might lead to the solutions with two angular momenta, and then we can compare the solution with the Myers-Perry solution with two angular momenta. \ We used the five-dimensional Minkowski metric as the seed metric in order to obtain soliton solutions. \ It would be interesting to study the solutions obtained by using non-Minkowski metric as the seed metric. \ It will be interesting to study the structure of the solutions with higher soliton numbers. \ It is possible to obtain these solutions by using the inverse scattering method. \ They can be regarded as the solutions expressing the interaction of two black holes, two black rings, or black hole and black ring.

\end{document}